\documentclass[aps,prl,twocolumn,superscriptaddress,10pt]{revtex4-2}
\usepackage[T1]{fontenc}
\usepackage[latin9]{inputenc}
\usepackage{geometry}
\usepackage{xcolor}
\usepackage{babel}
\usepackage{units}
\usepackage{amsmath}
\usepackage{amssymb}
\usepackage{stmaryrd}
\usepackage{graphicx}
\usepackage{wasysym}
\usepackage{bbold}
\usepackage[bookmarks=false]{hyperref}
\usepackage{orcidlink}
\makeatletter

\IfFileExists{lmodern.sty}{\usepackage{lmodern}}{}
\usepackage{babel}

\makeatother

\begin{document}
\title{Suppressing Intrinsic Spin-Phonon Errors in Trapped-Ion Quantum Simulation}
\author{Dylan Sheils\orcidlink{0000-0003-1276-7113}}
\thanks{These authors contributed equally to this work.}

\author{James Wang \orcidlink{0009-0009-5261-5879}}
\thanks{These authors contributed equally to this work.}

\author{Or Katz\orcidlink{0000-0001-7634-1993}}
\email{or.katz@cornell.edu}
\affiliation{School of Applied and Engineering Physics, Cornell University, Ithaca, NY 14853 USA}

\begin{abstract}
Trapped-ion quantum simulators realize programmable spin models through phonon-mediated interactions. For Hamiltonians with noncommuting terms, however, the same phonon bus generates intrinsic spin-phonon errors that strongly distort the target dynamics. Because these errors are governed by the full time history of the spin-dependent phonon motion, they survive standard loop-closing control and limit simulation accuracy. Using a sequence of frame transformations, we isolate the residual error dynamics and show that this intrinsic error can be strongly suppressed while preserving programmable Ising couplings. Full spin-boson simulations of multi-ion chains demonstrate orders-of-magnitude lower error than both constant-drive and conventional loop-closing protocols. These results remove a central precision barrier in trapped-ion analog quantum simulation and enable accurate programmable simulation of noncommuting many-body Hamiltonians and dynamical protocols.
\end{abstract}
\maketitle

Quantum simulators provide access to interacting many-body dynamics that are difficult to compute classically, but their accuracy is often limited by couplings to auxiliary degrees of freedom that mediate the interactions while perturbing the target Hamiltonian. Trapped ions are a leading platform in this direction, combining programmable phonon-mediated spin-spin couplings with precise control \cite{Monroe2021Programmable,Blatt2012Quantum,Porras2004Effective,Deng2005Effective,Friedenauer2008Simulating,Britton2012Engineered,Senko2015Realization,Lu2025Implementing,shapira2025programmable}. Many models of current interest, however, require noncommuting terms such as transverse fields, as in the transverse-field Ising model, where quantum fluctuations qualitatively reshape the many-body dynamics \cite{Edwards2010Quantum,Islam2011Onset,Li2023Probing,puebla2019kibble,Richerme2014NonLocal,Jurcevic2014Quasiparticle,Smith2016ManyBody,Zhang2017Observation,Kaplan2020ManyBody,de2024observation,schuckert2025observation,katz2025hybrid}. In this regime, the same phonon modes that generate the desired interactions also create an intrinsic source of simulation error that is not fully characterized by final-time phonon displacement alone \cite{wang2012intrinsic,wall2017boson}.

In the commuting limit relevant to digital entangling gates, it is sufficient to require that the spin-dependent phonon trajectories return to their starting point in phase space at the target time, so that the phonons disentangle from the spins \cite{molmer1999multiparticle,sorensen2000entanglement,hayes2012coherent,leung2018robust,shapira2023robust}. For analog simulation with noncommuting fields, however, residual spin-phonon coupling during the evolution generates additional errors even when all phase-space trajectories close, qualitatively distorting the target dynamics. Dynamical-decoupling methods do not generically remove this effect because they are designed to refocus external noise rather than coherent mediator dynamics \cite{morong2023engineering}. Likewise, pulse-shaping methods developed for high-fidelity entangling gates and robust trapped-ion control constrain only the endpoint of the phonon motion, not the history-dependent spin-phonon dressing that arises in noncommuting many-body dynamics \cite{hayes2012coherent,leung2018robust,shapira2020theory,shapira2023robust,milne2020phase,bermudez2012robust}, with other methods correcting for control errors but not intrinsic spin-phonon couplings \cite{steckmann2506error,jia2023angle}. As a result, intrinsic phonon-induced errors can accumulate over the long evolution times most relevant for analog quantum simulation, limiting fidelity precisely where quantum simulators should offer their greatest advantage.

Here we show that in the noncommuting regime the residual spin-phonon error surviving standard loop closure is governed by the time history of the spin-dependent displacement trajectories. Using a sequence of frame transformations, we isolate the residual error and derive experimentally compatible conditions that suppress it while preserving programmable Ising interactions. These conditions can be incorporated into efficient waveform optimization without classically simulating the target many-body dynamics. Full spin-boson simulations demonstrate orders-of-magnitude error reduction.

Analog trapped-ion quantum simulations use spin-dependent drives to couple effective spin-$\tfrac{1}{2}$ degrees of freedom, encoded in the ions' internal states, to the crystal's collective phonon modes. For near-symmetric bichromatic driving close to the first red and blue sidebands, in the Lamb-Dicke regime and within the rotating-wave approximation, the dynamics of $N$ ion spins are described by the spin-boson Hamiltonian \cite{Monroe2021Programmable,Porras2004Effective}
\begin{equation}\begin{aligned}\vspace{-3pt}
\hat{H}=&\sum_k \omega_k \hat{n}_k+\tfrac{1}{2}\sum_j B_j \hat{\sigma}_z^{(j)}+\sum_{j,k}\bigl(g_{jk}\,\hat{a}_k^\dagger+g_{jk}^*\,\hat{a}_k\bigr)\hat{\sigma}_x^{(j)}
\label{eq:H_drive}
\end{aligned}\vspace{-3pt}\end{equation}
with $\hbar=1$. Here $\hat{a}_k$ annihilates phonon mode $k$ of frequency $\omega_k$, $\hat{n}_k=\hat{a}_k^\dagger\hat{a}_k$ is the number operator, and $\hat{\sigma}_\mu^{(j)}$ are Pauli operators acting on ion $j$. The field $B_j$ acts along $\hat{\sigma}_z$, while the spin-dependent force couples through $\hat{\sigma}_x$. For bichromatic sideband driving, $g_{jk}(t)=\eta_k b_{jk}\Omega_j(t)e^{-i\nu t}/2$, where $\eta_k$ is the Lamb-Dicke factor, $b_{jk}$ is the participation of ion $j$ in mode $k$, $\Omega_j(t)$ is the complex Rabi-frequency envelope, and $\nu$ is the bichromatic beatnote, detuned from mode $k$ by $\delta_k=\omega_k-\nu$.

\begin{figure*}[t]
    \centering
    \includegraphics[width=1.8\columnwidth]{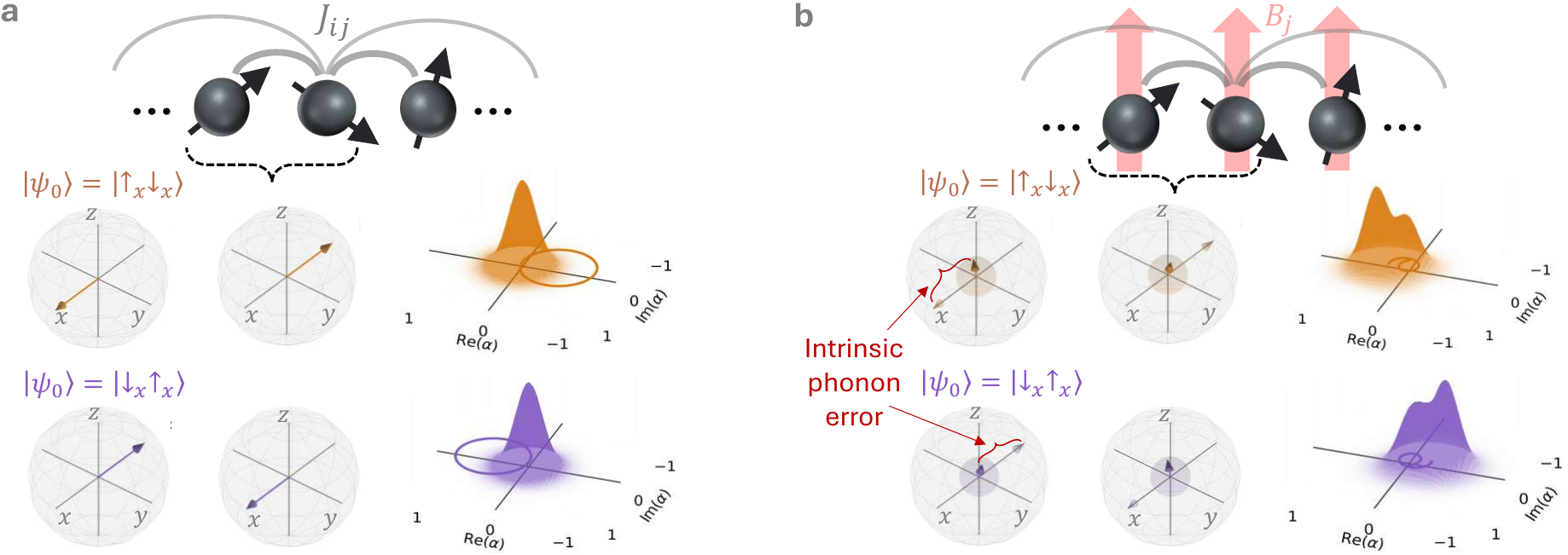}\vspace{-3pt}
    \caption{\textbf{Intrinsic phonon error in trapped-ion quantum simulation.} Full spin-boson simulations for a minimal two-spin, one-mode system. Spin-dependent forces couple the spins to the phonon mode that mediates the Ising interaction, generating trajectories in the complex displacement plane $\alpha$. \textbf{a}, Commuting limit ($B_j=0$). For representative $\hat{\sigma}_x$-basis states (orange and purple), the spin-conditioned trajectories close at the target time, and the phonon wave packet remains localized and Gaussian. \textbf{b}, Noncommuting regime ($B_j\neq 0$). For the same control fields, the transverse field makes the motion history dependent, generating residual spin-phonon correlations. To isolate this intrinsic fidelity-limiting error, we apply the ideal inverse evolution $U_{\mathrm{TFIM}}^\dagger$ to the simulated state. Tracing out the spins yields the reduced phonon state, shown as a Husimi function, whose structure is no longer Gaussian. Tracing out the phonon mode yields decohered reduced spin states (Bloch spheres), revealing the intrinsic phonon error that limits simulation fidelity. See \cite{SI} for simulation details.}\vspace{-12pt}
    \label{fig:overview}
\end{figure*}

To analyze the intrinsic simulation error, we move to the interaction picture of the free phonon evolution, in which the spin-dependent force acquires the phases $e^{\pm i\omega_k t}$. The corresponding interaction-picture propagator can then be written in Magnus form as
 \cite{Monroe2021Programmable,Porras2004Effective} \begin{equation}
\tilde{U}(T)=\exp\!\Bigl[-i\!\int_0^T \hat{H}_{\scriptscriptstyle \mathrm{TFIM}}(t)\,dt+\hat{\Omega}_{\rm err}(T)\Bigr],
\label{eq:U_TFIM_err}
\end{equation}
where $\hat{H}_{\scriptscriptstyle \mathrm{TFIM}}(t)=\sum_{i<j}J_{ij}(t)\hat{\sigma}_x^{(i)}\hat{\sigma}_x^{(j)}+\frac{1}{2}\sum_j B_j \hat{\sigma}_z^{(j)}$ is the target transverse-field Ising Hamiltonian with drive-generated couplings $J_{ij}(t)$, and $\hat{\Omega}_{\rm err}(T)$ collects the residual spin-phonon terms. This decomposition may be obtained, for example, from a Magnus expansion \cite{wang2012intrinsic}, although the control construction developed below does not rely on truncating that expansion.

The commuting limit $B_j=0$ recovers the standard intuition. In that case, $\hat{\Omega}_{\rm err}(T)=\sum_{j,k}\bigl[\alpha_{jk}(T)\hat{a}_k^\dagger-\alpha_{jk}^*(T)\hat{a}_k\bigr]\hat{\sigma}_x^{(j)}$, where $\alpha_{jk}(t)=-i\int_0^t g_{jk}(\tau)e^{i\omega_k\tau}\,d\tau$ is the spin-dependent phase-space displacement of mode $k$ associated with ion $j$. Final-time loop closure, $\alpha_{jk}(T)=0$, then removes the spin-phonon entanglement and recovers pure Ising evolution \cite{molmer1999multiparticle,sorensen2000entanglement,leung2018robust,lu2019global}. This is the familiar logic underlying high-fidelity entangling gates.

With noncommuting fields, however, this picture breaks down. The spin-dependent force in Eq.~\eqref{eq:H_drive}, which acts as a phonon-conditioned rotation about $\hat{\sigma}_x^{(j)}$, no longer commutes with the transverse-field term $B_j\hat{\sigma}_z^{(j)}$. Final-time loop closure therefore no longer suffices to control the error. In a Magnus description, nested commutators generate residual spin-phonon terms involving additional spin components and higher-order bosonic operators \cite{wang2012intrinsic,wall2017boson}. Crucially, the weights of these terms are governed by the low-order temporal moments of the displacement trajectories $\alpha_{jk}(t)$ and therefore need not vanish even when $\alpha_{jk}(T)=0$. For many spins and modes, this produces a rapidly proliferating hierarchy of residual spin-phonon correlations, allowing errors to accumulate over the long evolution times relevant for analog simulation. This difficulty is especially acute when $|B_j|\gtrsim |J_{ij}|$, where the transverse field is not a small perturbation of the target spin model. As illustrated in Fig.~\ref{fig:overview}, this mechanism can already produce large deviations in a minimal two-spin, one-mode example for controls that would be error-free at $B=0$. The leading terms and their scaling are derived in the Supplemental Material \cite{SI}. In the noncommuting regime, final-time phonon disentanglement alone does not guarantee accurate spin-model dynamics.

To control this history-dependent error, we seek a frame in which the target spin dynamics are explicit while the residual spin-phonon motion is as stationary as possible, without solving the full many-body evolution under $\hat{H}_{\scriptscriptstyle \mathrm{TFIM}}$. We focus on the off-resonant regime relevant for nontrivial transverse-field dynamics, where the characteristic scales satisfy $B\sim J$ and $B\ll g\ll\delta$, so that the displacement scale $|\alpha|\sim g/\delta\ll1$ provides a controlled small parameter. From the phonon perspective, the interaction-picture Hamiltonian describes forced harmonic oscillators whose instantaneous equilibrium positions depend on the spin configuration. This motivates a spin-dependent displacement into the moving equilibrium frame,
\begin{equation}\vspace{-4pt}
\hat{\mathcal D}(t)=\prod_{j=1}^{N} \exp\!\Bigl[\hat{\sigma}_x^{(j)}\sum_k\bigl(\alpha_{jk}(t)\hat{a}_k^\dagger-\alpha_{jk}^*(t)\hat{a}_k\bigr)\Bigr].
\label{eq:D_def}\vspace{-1pt}
\end{equation}
Under this transformation the Hamiltonian becomes $\hat{H}_{\mathcal D}(t)=\hat{H}_{\scriptscriptstyle \mathrm{TFIM}}(t)+\hat{\epsilon}(t)$, with
\begin{equation}
\hat{\epsilon}(t)=\sum_jB_j\left(\tfrac{1}{2}\hat{\sigma}_y^{(j)}\sin(\hat{\theta}_j)-\hat{\sigma}_z^{(j)}\sin^2(\tfrac{1}{2}\hat{\theta}_j)\right),\label{eq:H_D}\vspace{-7pt}\end{equation}
where $\hat{\theta}_j(t)=2i\sum_k\bigl(\alpha_{jk}(t)\hat{a}_k^\dagger-\alpha_{jk}^*(t)\hat{a}_k\bigr)$. In this frame the Ising couplings appear explicitly as $J_{ij}=-2\mathrm{Im}\bigl(\sum_k \alpha_{ik}^*\dot{\alpha}_{jk}\bigr)$, that is, through the phase-space area enclosed by the displacement trajectories. The residual term $\hat{\epsilon}(t)$ has a simple interpretation: it is a phonon-dependent rotation of spin $j$ about $\hat{\sigma}_x^{(j)}$ by the operator-valued angle $\hat{\theta}_j(t)$. Because the same displacements $\alpha_{jk}(t)$ generate both $J_{ij}(t)$ and $\hat{\theta}_j(t)$, the error cannot be eliminated identically; instead, it can be strongly suppressed by shaping its temporal structure.

A key observation is that $\hat{H}_{\scriptscriptstyle \mathrm{TFIM}}(t)$ and $\hat{\epsilon}(t)$ respond differently to the temporal structure of $\alpha_{jk}(t)$. We therefore choose the drives so that each displacement trajectory has vanishing low-order temporal moments,
\begin{equation}
\int_0^T dt\, \alpha_{jk}(t)\,P_{n}\!\left(\frac{2t}{T}-1\right)=0,
\label{eq:low_freq_alpha}
\end{equation}
using shifted-Legendre polynomials $P_{n}$ as a convenient experimentally compatible filter set \cite{SI}. These constraints suppress the low-frequency (secular) content of $\alpha_{jk}(t)$ and push the displacement trajectories toward motion concentrated near the detuning scale. The Ising interaction, however, depends on the bilinear combination $\alpha_{ik}^*(t)\dot{\alpha}_{jk}(t)$ rather than on $\alpha_{jk}(t)$ itself. Consequently, even when $\alpha_{jk}(t)$ carries little or no secular weight, the product $\alpha_{ik}^*(t)\dot{\alpha}_{jk}(t)$ can retain the component needed to generate the desired spin-spin coupling.

\begin{figure}[t]
    \centering
    \includegraphics[width=1\columnwidth]{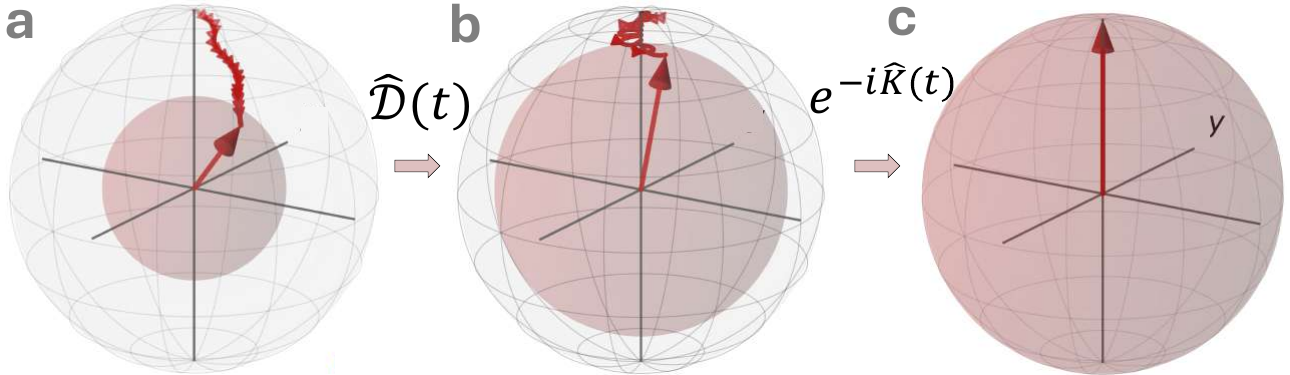}\vspace{-3pt}
    \caption{\textbf{Strategy for suppressing residual spin-phonon error.} Schematic reduced Bloch-sphere trajectories for rapidly modulated $\alpha_{jk}(t)$. \textbf{a}, In the drive-defined frame (Eq.~\ref{eq:H_drive}), residual spin-phonon correlations make the reduced spin state mixed, so the Bloch vector lies inside the sphere. \textbf{b}, A spin-dependent displacement $\hat{\mathcal D}(t)$ (Eq.~\ref{eq:D_def}) moves to the instantaneous phonon-equilibrium frame, removing most of the phonon dressing but leaving the transverse-field-induced phonon-conditioned micromotion. \textbf{c}, A further kick transformation $e^{-i\hat K(t)}$ (Eq.~\ref{eq:Kick_operator}) removes this remaining micromotion, so the reduced spin trajectory nearly recovers the ideal slow target dynamics.}\vspace{-12pt}
    \label{fig:solution}
\end{figure}

Equation~\eqref{eq:H_D} separates into a fast oscillatory part and a secular part. The latter, denoted by $\hat{\epsilon}_0$, is a phonon-dependent renormalization of the effective transverse field. Its exact form is given in the Supplemental Material \cite{SI}; in the Fock basis it is expressed in terms of Laguerre polynomials, and thermal averaging yields the familiar Debye-Waller factor \cite{leibfried2003quantum}. To lowest order in $|\alpha|^2$,
\begin{equation}
\hat{\epsilon}_0 \approx -\sum_{j,k} B_j \hat{\sigma}_z^{(j)}|\alpha_{jk}|^2(2\hat{n}_k+1).
\label{eq:epsilon_0}
\end{equation}
The phonon-independent part yields a known spin-only renormalization that reduces the effective transverse field and can be absorbed into a re-calibrated target Hamiltonian, whereas the occupation-dependent part is the intrinsic residual error for thermally populated modes \cite{wall2017boson}.

The remaining fast component of $\hat{\epsilon}(t)$ acts as a rapid phonon-conditioned rotation on the Bloch sphere, reminiscent of micromotion in a Floquet problem. This motivates introducing a kick operator \cite{eckardt2015high,rahav2003effective}, here without requiring periodicity. A further unitary transformation $e^{-i\hat{K}(t)}$ generated by
\begin{equation}
\hat{K}(t)=\int_0^t d\tau\,\bigl[\hat{\epsilon}(\tau)-\hat{\epsilon}_0(\tau)\bigr],
\label{eq:Kick_operator}
\end{equation}
defines a kick frame in which the Hamiltonian becomes $\hat{H}_K(t)=\hat{H}_{\scriptscriptstyle \mathrm{TFIM}}(t)+\hat{\epsilon}_0(t)+\hat{\epsilon}_K(t)$, where $\hat{\epsilon}_K(t)$ arises from the back-action of the slow TFIM dynamics on the fast motion. Its exact form is derived in the Supplemental Material \cite{SI} and is parametrically suppressed, with characteristic scale $\epsilon_K\sim \max(B,J)\,|\alpha|(B/\delta)$. The kick transformation therefore isolates the fast component, leaving the residual error controlled by the secular term $\hat{\epsilon}_0(t)$ and the parametrically smaller correction $\hat{\epsilon}_K(t)$.

Experiments, however, probe the dynamics in the drive-defined rotating frame of Eq.~\eqref{eq:H_drive}, rather than in the kick frame. Undoing the successive frame transformations gives the corresponding time-evolution operator in that frame, \begin{equation}
\hat{U}(T)=\hat{U}_{\rm ph}(T)\hat{\mathcal D}(T)e^{-i\hat{K}(T)}
\exp\!\Bigl[-i\!\int_0^T \hat{H}_K\,dt\Bigr],
\label{eq:U_lab}
\end{equation}
where we assume $\alpha_{jk}(0)=0$, so that $\hat{\mathcal D}(0)=e^{-i\hat{K}(0)}=\hat{\mathbb I}$. Requiring closure of both the displacement and the kick operator at the target time,
\begin{equation}
\alpha_{jk}(T)=0,
\qquad
\hat{K}(T)=0,
\label{eq:constraints2}
\end{equation}
gives $\hat{\mathcal D}(T)=e^{-i\hat{K}(T)}=\hat{\mathbb I}$, removing both the phase-space displacement and the kick-frame mismatch. The evolution in the frame of Eq.~\eqref{eq:H_drive} then coincides with that generated by $\hat H_K(t)$, up to the free phonon evolution unitary $\hat U_{\rm ph}(T)$; the resulting spin dynamics, and hence spin observables, are therefore those generated by $\hat H_K(t)$. Although $\hat{K}(T)$ contains a hierarchy of higher-order, history-dependent phonon operators, to leading order in $\alpha$ the condition $\hat{K}(T)=0$ reduces to $\int_0^T dt\,\alpha_{jk}(t)=0$, while our optimization cancels higher-order contributions as well, to controllable accuracy \cite{SI}. Figure~\ref{fig:solution} summarizes the sequence of frame transformations and the resulting control conditions.

\begin{figure}[t]
    \centering
    \includegraphics[width=1\columnwidth]{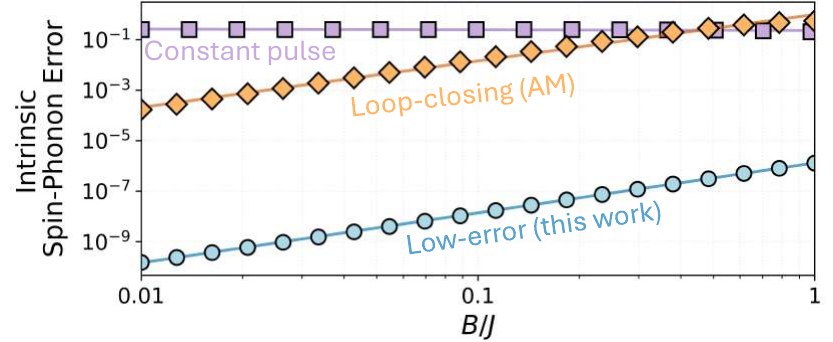}\vspace{-5pt}
    \caption{\textbf{Intrinsic spin-phonon error benchmark.} Full spin-boson simulations for a four-ion chain with four ground-state-cooled phonon modes. The intrinsic error is obtained by removing the intended TFIM evolution, tracing out the phonons, and quantifying the resulting nonunitarity of the reduced spin channel (see \cite{SI}). Across a broad range of $B$, the optimized low-error waveform of this work (blue circles) suppresses the error by orders of magnitude relative to both a constant pulse (purple squares) and an amplitude-modulated pulse constrained only by loop closure, $\alpha_{jk}(T)=0$ (orange diamonds).}\vspace{-10pt}
    \label{fig:benchmarks}
\end{figure}

Under the closure conditions in Eq.~\eqref{eq:constraints2} and the temporal-moment constraints in Eq.~\eqref{eq:low_freq_alpha}, the implemented spin dynamics are governed by $\hat{H}_K(t)$ up to the known phonon-only factor $\hat{U}_{\rm ph}(T)$. The residual terms therefore directly quantify the error of the implemented spin dynamics relative to the target TFIM. In the controlled short-time regime $T\lesssim 1/\max(B,J)$, where the residual Magnus expansion remains valid, the average infidelity is bounded by
\begin{equation}
1-\mathcal{F}\lesssim (\epsilon_0T)^2+(\epsilon_KT)^2.
\end{equation}
Here $\epsilon_0$ is the characteristic scale of the occupation-dependent part of $\hat{\epsilon}_0$. For thermal phonons with characteristic standard deviation $\Delta n$, one has $\epsilon_0\sim B|\alpha|^2\Delta n$. Taking a controlled comparison time $T\sim 1/B$ and representative parameters $B/g\sim0.1$, $g/\delta\sim|\alpha|\sim0.1$, and $\Delta n\lesssim 0.1$, we obtain $1-\mathcal{F}\lesssim10^{-6}$. By contrast, the standard short-time error in the noncommuting regime already scales as $(|\alpha|BT)^2$, which at the same comparison time is of order $10^{-2}$. This estimate is intended as a controlled short-time metric for intrinsic error generation and for comparison to uncontrolled waveforms; full spin-boson simulations confirm that the same waveforms maintain orders-of-magnitude error suppression well beyond this regime (see Fig.~\ref{fig:benchmarks} and \cite{SI}). Derivation of those estimates and higher-order corrections to the kick operator that allow further improvement are described in the Supplemental Material \cite{SI}.

To incorporate these control conditions into Hamiltonian engineering, we parameterize each drive envelope $\Omega_j(t)$ in a narrow-bandwidth basis of $M$ sinusoidal components with coefficients $c_{jm}$, allowing simultaneous amplitude and phase modulation. To leading order in $|\alpha|$, the temporal-moment and closure conditions define a set of linear constraints on these coefficients. The null space of the resulting constraint matrix provides an admissible family of waveforms that satisfy the low-error conditions at leading order. We then use this waveform family as the control space for a nonlinear optimization that targets the desired coupling elements $\int_0^T J_{ij}(t)\,dt$ that determine the evolution operator while further suppressing $\hat{K}(T)$ beyond leading order in $|\alpha|$. Further details of the parametrization and optimization are given in the Supplemental Material \cite{SI}.

\begin{figure}[b]
    \centering
    \includegraphics[width=1\columnwidth]{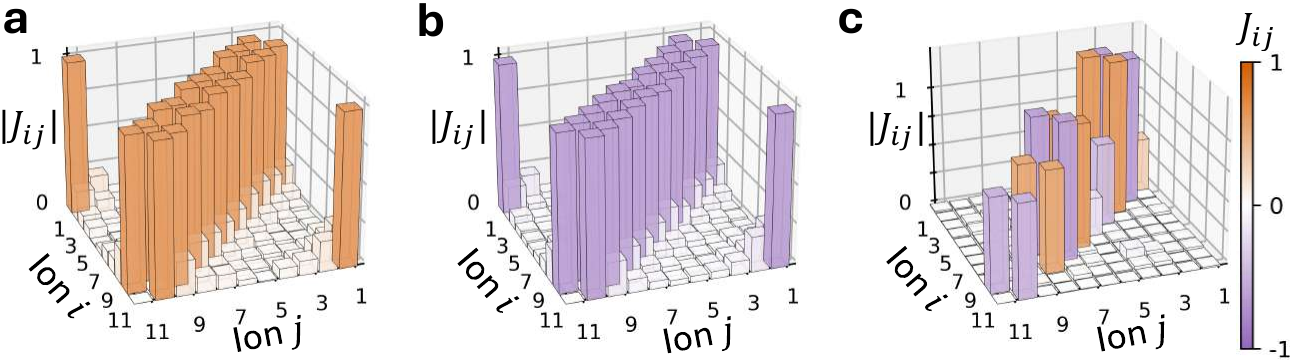}\vspace{-10pt}
    \caption{\textbf{Programmable interaction graphs under low-error constraints.} Optimized low-error realizations of representative Ising matrices $J_{ij}$ in an $11$-ion chain, showing that broad interaction programmability is retained under the error-suppressing constraints. Shown are long-range couplings with effective periodic boundary conditions in a linear chain (\textbf{a}), sign-reversed couplings enabling time-reversal protocols (\textbf{b}), and oscillatory quasiperiodic couplings (\textbf{c}). Bar height denotes $|J_{ij}|$, and color denotes the sign of $J_{ij}$.}\vspace{-7pt}
    \label{fig:programmability}
\end{figure}

To benchmark this procedure, we perform full spin-boson simulations for a four-ion chain, restricting attention to the four collective motional modes addressed by a spin-dependent drive aligned along one trap axis. Our optimized low-error waveform is compared against two standard control families designed to realize the same target interaction graph: a constant pulse, representative of typical analog-simulation drives \cite{Monroe2021Programmable,Porras2004Effective,Islam2011Onset,Britton2012Engineered}, and an amplitude-modulated pulse constrained only by loop closure, $\alpha_{jk}(T)=0$, which reaches high fidelity in the commuting limit $B=0$ \cite{hayes2012coherent,leung2018robust,shapira2023robust,milne2020phase}. All control families are compared at the same total duration $T=\pi/(4J)$ and under the same peak drive-amplitude constraint. The intrinsic spin-phonon error is quantified by removing the ideal target TFIM evolution, tracing out the phonons, and measuring the nonunitarity of the reduced spin channel. Since a purely spin-only mismatch remains unitary after this subtraction, the resulting nonunitarity isolates the irreversible error generated by residual history-dependent spin-phonon entanglement~\cite{SI}. As shown in Fig.~\ref{fig:benchmarks} for ground-state-cooled phonon modes, the optimized waveform maintains low error across a broad range of $B/J$, whereas the comparison protocols degrade rapidly once the transverse field is no longer perturbative. In the Supplemental Material we also demonstrate favorable behavior for finite thermal phonon states with mean occupation $\bar{n}$, where the remaining error is dominated by the secular contribution $\hat{\epsilon}_0$, consistent with the analysis above. These benchmarks validate that temporal-moment filtering together with displacement and kick closure yields orders-of-magnitude lower intrinsic error than both constant and loop-closing controls in the regime relevant for transverse-field dynamics.

Our waveform optimization, together with the low-error conditions in Eqs.~\eqref{eq:low_freq_alpha} and \eqref{eq:constraints2}, also preserves broad programmability of the engineered interactions \cite{Lu2025Implementing,shapira2025programmable,katz2023programmable_nbody,katz2023threefourbody,katz2023bosonic,feng2023continuous,katz2025floquet}. Figure~\ref{fig:programmability} shows optimized low-error solutions for several target matrices $J_{ij}$ in an $11$-ion chain, including short-range modulated interactions, long-range interactions with periodic boundary conditions, and sign-reversed interactions, enabling frustrated, quasiperiodic, scrambling, and time-reversal protocols, as well as broader nonequilibrium dynamical studies \cite{crowley2018quasiperiodic,chandran2017localization,Gaerttner2017Measuring,Lin2011Sharp,bermudez2017longrange,dumitrescu2022dynamical,morong2021stark,katz2025floquet}. 

The classical preprocessing required by our method solves a constrained control-design problem, rather than the full many-body quantum dynamics generated by the target Hamiltonian. In particular, the low-error conditions enter primarily as linear constraints on the waveform coefficients, while the remaining optimization targets the desired interaction matrix and residual kick-frame corrections~\cite{SI}. This procedure is computationally efficient and can be performed on most experimental platforms using standard narrow-bandwidth modulators. Consequently, the many-body dynamics of the target noncommuting spin model are generated by the quantum simulator itself, rather than by a classical simulation used to design the controls.

The resulting controls require only narrow-bandwidth amplitude and phase modulation and are compatible with programmable interaction engineering in trapped-ion platforms \cite{Lu2025Implementing,shapira2025programmable,Guo2024SiteResolved,wu2025qubits,Qiao2024Tunable,karpa2016phasestable}. By deriving constructive, history-dependent control conditions and validating them in full spin-boson simulations, our approach extends trapped-ion analog simulation to a broader class of noncommuting spin models. The same control principle may extend to other bosonic-mediator architectures \cite{barreiro2011opensystem,davoudi2020towards,safavi2018verification,ge2019squeezed,sutherland2021universal,katz2022nbody,katz2023bosonic} and to noncommuting Hamiltonian protocols outside spin simulation whenever the intrinsic error admits an analogous frame-separated description.
\bibliography{Refs}

@article{Monroe2021Programmable,
  author  = {Christopher Monroe and W. C. Campbell and L.-M. Duan and Zhe-Xuan Gong and Alexey V. Gorshkov and Paul Hess and Rajibul Islam and Kihwan Kim and Norbert M. Linke and Guido Pagano and Philip Richerme and Crystal Senko and Norman Y. Yao},
  title   = {Programmable quantum simulations of spin systems with trapped ions},
  journal = {Reviews of Modern Physics},
  volume  = {93},
  number  = {2},
  pages   = {025001},
  year    = {2021},
  doi = {10.1103/RevModPhys.93.025001}
}

@article{Blatt2012Quantum,
  author  = {Rainer Blatt and Christian F. Roos},
  title   = {Quantum simulations with trapped ions},
  journal = {Nature Physics},
  volume  = {8},
  number  = {4},
  pages   = {277--284},
  year    = {2012},
  doi = {10.1038/nphys2252}
}

@article{Porras2004Effective,
  author  = {Diego Porras and J. Ignacio Cirac},
  title   = {Effective quantum spin systems with trapped ions},
  journal = {Physical Review Letters},
  volume  = {92},
  number  = {20},
  pages   = {207901},
  year    = {2004},
  doi = {10.1103/PhysRevLett.92.207901}
}

@article{Deng2005Effective,
  author  = {Xiao-Liang Deng and Diego Porras and J. Ignacio Cirac},
  title   = {Effective spin quantum phases in systems of trapped ions},
  journal = {Physical Review A},
  volume  = {72},
  number  = {6},
  pages   = {063407},
  year    = {2005},
  doi = {10.1103/PhysRevA.72.063407}
}

@article{Friedenauer2008Simulating,
  author  = {A. Friedenauer and H. Schmitz and J. T. Glueckert and D. Porras and T. Schaetz},
  title   = {Simulating a quantum magnet with trapped ions},
  journal = {Nature Physics},
  volume  = {4},
  number  = {10},
  pages   = {757--761},
  year    = {2008},
  doi = {10.1038/nphys1032}
}

@article{Edwards2010Quantum,
  author  = {E. E. Edwards and S. Korenblit and Kihwan Kim and Rajibul Islam and M.-S. Chang and J. K. Freericks and G.-D. Lin and L.-M. Duan and Christopher Monroe},
  title   = {Quantum simulation and phase diagram of the transverse-field Ising model with three atomic spins},
  journal = {Physical Review B},
  volume  = {82},
  number  = {6},
  pages   = {060412(R)},
  year    = {2010},
  doi = {10.1103/PhysRevB.82.060412}
}

@article{Islam2011Onset,
  author  = {Rajibul Islam and E. E. Edwards and Kihwan Kim and S. Korenblit and C. Noh and H. Carmichael and G.-D. Lin and L.-M. Duan and C.-C. Joseph Wang and J. K. Freericks and Christopher Monroe},
  title   = {Onset of a quantum phase transition with a trapped ion quantum simulator},
  journal = {Nature Communications},
  volume  = {2},
  pages   = {377},
  year    = {2011},
  doi = {10.1038/ncomms1374}
}

@article{Lin2011Sharp,
  author  = {G.-D. Lin and Christopher Monroe and L.-M. Duan},
  title   = {Sharp Phase Transitions in a Small Frustrated Network of Trapped Ion Spins},
  journal = {Physical Review Letters},
  volume  = {106},
  number  = {23},
  pages   = {230402},
  year    = {2011},
  doi = {10.1103/PhysRevLett.106.230402}
}

@article{Britton2012Engineered,
  author  = {Joseph W. Britton and Brian C. Sawyer and Adam C. Keith and C.-C. Joseph Wang and James K. Freericks and Hermann Uys and John J. Bollinger},
  title   = {Engineered two-dimensional Ising interactions in a trapped-ion quantum simulator with hundreds of spins},
  journal = {Nature},
  volume  = {484},
  number  = {7395},
  pages   = {489--492},
  year    = {2012},
  doi = {10.1038/nature10981}
}

@article{Richerme2014NonLocal,
  author  = {Philip Richerme and Zhe-Xuan Gong and Aaron Lee and Crystal Senko and Jacob Smith and Michael Foss-Feig and Spyridon Michalakis and Alexey V. Gorshkov and Christopher Monroe},
  title   = {Non-local propagation of correlations in quantum systems with long-range interactions},
  journal = {Nature},
  volume  = {511},
  number  = {7508},
  pages   = {198--201},
  year    = {2014},
  doi = {10.1038/nature13450}
}

@article{Jurcevic2014Quasiparticle,
  author  = {P. Jurcevic and B. P. Lanyon and P. Hauke and C. Hempel and P. Zoller and R. Blatt and C. F. Roos},
  title   = {Quasiparticle engineering and entanglement propagation in a quantum many-body system},
  journal = {Nature},
  volume  = {511},
  number  = {7508},
  pages   = {202--205},
  year    = {2014},
  doi = {10.1038/nature13461}
}

@article{Senko2015Realization,
  author  = {Crystal Senko and Philip Richerme and Jacob Smith and Aaron Lee and I. Cohen and A. Retzker and Christopher Monroe},
  title   = {Realization of a quantum integer-spin chain with controllable interactions},
  journal = {Physical Review X},
  volume  = {5},
  number  = {2},
  pages   = {021026},
  year    = {2015},
  doi = {10.1103/PhysRevX.5.021026}
}

@article{Smith2016ManyBody,
  author  = {J. Smith and A. Lee and P. Richerme and B. Neyenhuis and P. W. Hess and P. Hauke and M. Heyl and D. A. Huse and C. Monroe},
  title   = {Many-body localization in a quantum simulator with programmable random disorder},
  journal = {Nature Physics},
  volume  = {12},
  number  = {10},
  pages   = {907--911},
  year    = {2016},
  doi = {10.1038/nphys3783}
}

@article{Zhang2017Observation,
  author  = {J. Zhang and G. Pagano and P. W. Hess and A. Kyprianidis and P. Becker and H. Kaplan and A. V. Gorshkov and Z.-X. Gong and C. Monroe},
  title   = {Observation of a Many-Body Dynamical Phase Transition with a 53-Qubit Quantum Simulator},
  journal = {Nature},
  volume  = {551},
  number  = {7682},
  pages   = {601--604},
  year    = {2017},
  doi = {10.1038/nature24654}
}

@article{Gaerttner2017Measuring,
  author  = {Martin G{\"a}rttner and Justin G. Bohnet and Arghavan Safavi-Naini and Michael L. Wall and John J. Bollinger and Ana Maria Rey},
  title   = {Measuring out-of-time-order correlations and multiple quantum spectra in a trapped-ion quantum magnet},
  journal = {Nature Physics},
  volume  = {13},
  number  = {8},
  pages   = {781--786},
  year    = {2017},
  doi = {10.1038/nphys4119}
}

@article{Kaplan2020ManyBody,
  author  = {Harvey B. Kaplan and Lingzhen Guo and Wen Lin Tan and Arinjoy De and Florian Marquardt and Guido Pagano and Christopher Monroe},
  title   = {Many-Body Dephasing in a Trapped-Ion Quantum Simulator},
  journal = {Physical Review Letters},
  volume  = {125},
  number  = {12},
  pages   = {120605},
  year    = {2020},
  doi = {10.1103/PhysRevLett.125.120605}
}

@article{Li2023Probing,
  author  = {B.-W. Li and Y.-K. Wu and Q.-X. Mei and R. Yao and W.-Q. Lian and M.-L. Cai and Y. Wang and B.-X. Qi and L. Yao and L. He and Z.-C. Zhou and L.-M. Duan},
  title   = {Probing Critical Behavior of Long-Range Transverse-Field Ising Model through Quantum Kibble-Zurek Mechanism},
  journal = {PRX Quantum},
  volume  = {4},
  number  = {1},
  pages   = {010302},
  year    = {2023},
  doi = {10.1103/PRXQuantum.4.010302}
}

@article{Qiao2024Tunable,
  author  = {Mu Qiao and Zhengyang Cai and Ye Wang and Botao Du and Naijun Jin and Wentao Chen and Pengfei Wang and Chunyang Luan and Erfu Gao and Ximo Sun and Haonan Tian and Jingning Zhang and Kihwan Kim},
  title   = {Tunable quantum simulation of spin models with a two-dimensional ion crystal},
  journal = {Nature Physics},
  volume  = {20},
  number  = {4},
  pages   = {642--648},
  year    = {2024},
  doi = {10.1038/s41567-023-02378-9}
}

@article{Guo2024SiteResolved,
  author  = {S.-A. Guo and Y.-K. Wu and J. Ye and L. Zhang and W.-Q. Lian and R. Yao and Y. Wang and R.-Y. Yan and Y.-J. Yi and Y.-L. Xu and B.-W. Li and Y.-H. Hou and Y.-Z. Xu and W.-X. Guo and C. Zhang and B.-X. Qi and Z.-C. Zhou and L. He and L.-M. Duan},
  title   = {A Site-Resolved 2D Quantum Simulator with Hundreds of Trapped Ions},
  journal = {Nature},
  volume  = {629},
  number  = {8014},
  pages   = {951--957},
  year    = {2024},
  doi = {10.1038/s41586-024-07459-0}
}

@article{Lu2025Implementing,
  author  = {Yao Lu and Wentao Chen and Shuaining Zhang and Kuan Zhang and Jialiang Zhang and Jing-Ning Zhang and Kihwan Kim},
  title   = {Implementing Arbitrary Ising Models with a Trapped-Ion Quantum Processor},
  journal = {Physical Review Letters},
  volume  = {134},
  number  = {5},
  pages   = {050602},
  year    = {2025},
  doi = {10.1103/PhysRevLett.134.050602}
}

@article{shapira2025programmable,
  title={Programmable Quantum Simulations on a Trapped-Ion Quantum Computer with a Global Drive},
  author={Shapira, Yotam and Markov, Jovan and Akerman, Nitzan and Stern, Ady and Ozeri, Roee},
  journal={Physical Review Letters},
  volume={134},
  pages={010602},
  year={2025},
  doi={10.1103/PhysRevLett.134.010602}
}

@article{wu2025qubits,
  title={Qubits on Programmable Geometries with a Trapped-Ion Quantum Processor},
  author={Wu, Qiming and Shi, Yue and Zhang, Jiehang},
  journal={Physical Review A},
  volume={111},
  pages={042607},
  year={2025},
  doi={10.1103/PhysRevA.111.042607}
}

@article{wang2012intrinsic,
  title={Intrinsic Phonon Effects on Analog Quantum Simulators with Ultracold Trapped Ions},
  author={Wang, C.-C. Joseph and Freericks, J. K.},
  journal={Physical Review A},
  volume={86},
  pages={032329},
  year={2012},
  doi={10.1103/PhysRevA.86.032329}
}

@article{wall2017boson,
  title={Boson-Mediated Quantum Spin Simulators in Transverse Fields: {XY} Model and Spin-Boson Entanglement},
  author={Wall, Michael L. and Safavi-Naini, Arghavan and Rey, Ana Maria},
  journal={Physical Review A},
  volume={95},
  pages={013602},
  year={2017},
  doi={10.1103/PhysRevA.95.013602}
}

@article{safavi2018verification,
  title={Verification of a Many-Ion Simulator of the Dicke Model Through Slow Quenches Across a Phase Transition},
  author={Safavi-Naini, A. and Lewis-Swan, R. J. and Bohnet, J. G. and G{\"a}rttner, M. and Gilmore, K. A. and Jordan, J. E. and Cohn, J. and Freericks, J. K. and Rey, A. M. and Bollinger, J. J.},
  journal={Physical Review Letters},
  volume={121},
  pages={040503},
  year={2018},
  doi={10.1103/PhysRevLett.121.040503}
}

@article{ge2019squeezed,
  title={Trapped-Ion Quantum Information Processing with Squeezed Phonons},
  author={Ge, Wenchao and Sawyer, Brian and Britton, Joe and Jacobs, Kurt and Bollinger, John and Foss-Feig, Michael},
  journal={Physical Review Letters},
  volume={122},
  pages={030501},
  year={2019},
  doi={10.1103/PhysRevLett.122.030501}
}

@article{puebla2019kibble,
  title={Quantum Kibble-Zurek Physics in Long-Range Transverse-Field Ising Models},
  author={Puebla, Ricardo and Marty, Oliver and Plenio, Martin B.},
  journal={Physical Review A},
  volume={100},
  pages={032115},
  year={2019},
  doi={10.1103/PhysRevA.100.032115}
}

@article{molmer1999multiparticle,
  title={Multiparticle Entanglement of Hot Trapped Ions},
  author={M{\o}lmer, Klaus and S{\o}rensen, Anders},
  journal={Physical Review Letters},
  volume={82},
  pages={1835--1838},
  year={1999},
  doi={10.1103/PhysRevLett.82.1835}
}

@article{sorensen2000entanglement,
  title={Entanglement and Quantum Computation with Ions in Thermal Motion},
  author={S{\o}rensen, Anders and M{\o}lmer, Klaus},
  journal={Physical Review A},
  volume={62},
  pages={022311},
  year={2000},
  doi={10.1103/PhysRevA.62.022311}
}

@article{hayes2012coherent,
  title={Coherent Error Suppression in Multiqubit Entangling Gates},
  author={Hayes, D. and Clark, S. M. and Debnath, S. and Hucul, D. and Inlek, I. V. and Lee, K. W. and Quraishi, Q. and Monroe, C.},
  journal={Physical Review Letters},
  volume={109},
  pages={020503},
  year={2012},
  doi={10.1103/PhysRevLett.109.020503}
}

@article{leung2018robust,
  title={Robust 2-Qubit Gates in a Linear Ion Crystal Using a Frequency-Modulated Driving Force},
  author={Leung, Pak Hong and Landsman, Kevin A. and Figgatt, Caroline and Linke, Norbert M. and Monroe, Christopher and Brown, Kenneth R.},
  journal={Physical Review Letters},
  volume={120},
  pages={020501},
  year={2018},
  doi={10.1103/PhysRevLett.120.020501}
}

@article{milne2020phase,
  title={Phase-Modulated Entangling Gates Robust to Static and Time-Varying Errors},
  author={Milne, Alistair R. and Edmunds, Claire L. and Hempel, Cornelius and Roy, Federico and Mavadia, Sandeep and Biercuk, Michael J.},
  journal={Physical Review Applied},
  volume={13},
  pages={024022},
  year={2020},
  doi={10.1103/PhysRevApplied.13.024022}
}

@article{morong2023engineering,
  title={Engineering Dynamically Decoupled Quantum Simulations with Trapped Ions},
  author={Morong, W. and Collins, K. S. and De, A. and Stavropoulos, E. and You, T. and Monroe, C.},
  journal={PRX Quantum},
  volume={4},
  pages={010334},
  year={2023},
  doi={10.1103/PRXQuantum.4.010334}
}

@article{shapira2020theory,
  title={Theory of Robust Multiqubit Nonadiabatic Gates for Trapped Ions},
  author={Shapira, Yotam and Shaniv, Ravid and Manovitz, Tom and Akerman, Nitzan and Peleg, Lee and Gazit, Lior and Ozeri, Roee and Stern, Ady},
  journal={Physical Review A},
  volume={101},
  pages={032330},
  year={2020},
  doi={10.1103/PhysRevA.101.032330}
}

@article{shapira2023robust,
  title={Robust Two-Qubit Gates for Trapped Ions Using Spin-Dependent Squeezing},
  author={Shapira, Yotam and Cohen, Sapir and Akerman, Nitzan and Stern, Ady and Ozeri, Roee},
  journal={Physical Review Letters},
  volume={130},
  pages={030602},
  year={2023},
  doi={10.1103/PhysRevLett.130.030602}
}

@article{bermudez2012robust,
  title={Robust Trapped-Ion Quantum Logic Gates by Continuous Dynamical Decoupling},
  author={Bermudez, A. and Schmidt, P. O. and Plenio, M. B. and Retzker, A.},
  journal={Physical Review A},
  volume={85},
  pages={040302},
  year={2012},
  doi={10.1103/PhysRevA.85.040302}
}

@article{bermudez2017longrange,
  title={Long-Range Heisenberg Models in Quasiperiodically Driven Crystals of Trapped Ions},
  author={Bermudez, A. and Tagliacozzo, L. and Sierra, G. and Richerme, P.},
  journal={Physical Review B},
  volume={95},
  pages={024431},
  year={2017},
  doi={10.1103/PhysRevB.95.024431}
}

@article{davoudi2020towards,
  title={Towards Analog Quantum Simulations of Lattice Gauge Theories with Trapped Ions},
  author={Davoudi, Zohreh and Hafezi, Mohammad and Monroe, Christopher and Pagano, Guido and Seif, Alireza and Shaw, Andrew},
  journal={Physical Review Research},
  volume={2},
  pages={023015},
  year={2020},
  doi={10.1103/PhysRevResearch.2.023015}
}

@article{schuckert2025observation,
  title={Observation of a finite-energy phase transition in a one-dimensional quantum simulator},
  author={Schuckert, Alexander and Katz, Or and Feng, Lei and Crane, Eleanor and De, Arinjoy and Hafezi, Mohammad and Gorshkov, Alexey V and Monroe, Christopher},
  journal={Nature Physics},
  volume={21},
  number={3},
  pages={374--379},
  year={2025},
  doi={10.1038/s41567-024-02751-2}
}

@article{barreiro2011opensystem,
  title={An Open-System Quantum Simulator with Trapped Ions},
  author={Barreiro, J. T. and M{\"u}ller, M. and Schindler, P. and Nigg, D. and Monz, T. and Chwalla, M. and Hennrich, M. and Roos, C. F. and Zoller, P. and Blatt, R.},
  journal={Nature},
  volume={470},
  pages={486--491},
  year={2011},
  doi={10.1038/nature09801}
}

@article{dumitrescu2022dynamical,
  title={Dynamical Topological Phase Realized in a Trapped-Ion Quantum Simulator},
  author={Dumitrescu, Philipp T. and Bohnet, Justin G. and Gaebler, John P. and Hankin, Aaron and Hayes, David and Kumar, Ajesh and Neyenhuis, Brian and Vasseur, Romain and Potter, Andrew C.},
  journal={Nature},
  volume={607},
  pages={463--467},
  year={2022},
  doi={10.1038/s41586-022-04853-4}
}

@article{morong2021stark,
  title={Observation of Stark Many-Body Localization Without Disorder},
  author={Morong, W. and Liu, F. and Becker, P. and Collins, K. S. and Feng, L. and Kyprianidis, A. and Pagano, G. and You, T. and Gorshkov, A. V. and Monroe, C.},
  journal={Nature},
  volume={599},
  pages={393--398},
  year={2021},
  doi={10.1038/s41586-021-03988-0}
}

@article{katz2022nbody,
  title={{N}-Body Interactions between Trapped Ion Qubits via Spin-Dependent Squeezing},
  author={Katz, Or and Cetina, Marko and Monroe, Christopher},
  journal={Physical Review Letters},
  volume={129},
  pages={063603},
  year={2022},
  doi={10.1103/PhysRevLett.129.063603}
}

@article{katz2023programmable_nbody,
  title={Programmable {N}-Body Interactions with Trapped Ions},
  author={Katz, Or and Cetina, Marko and Monroe, Christopher},
  journal={PRX Quantum},
  volume={4},
  pages={030311},
  year={2023},
  doi={10.1103/PRXQuantum.4.030311}
}

@article{katz2023threefourbody,
  title={Demonstration of Three- and Four-Body Interactions between Trapped-Ion Spins},
  author={Katz, Or and Feng, Lei and Risinger, Andrew and Monroe, Christopher and Cetina, Marko},
  journal={Nature Physics},
  volume={19},
  pages={1452--1458},
  year={2023},
  doi={10.1038/s41567-023-02102-7}
}

@article{katz2023bosonic,
  title={Programmable Quantum Simulations of Bosonic Systems with Trapped Ions},
  author={Katz, Or and Monroe, Christopher},
  journal={Physical Review Letters},
  volume={131},
  pages={033604},
  year={2023},
  doi={10.1103/PhysRevLett.131.033604}
}

@article{feng2023continuous,
  title={Continuous Symmetry Breaking in a Trapped-Ion Spin Chain},
  author={Feng, Lei and Katz, Or and Haack, Casey and Maghrebi, Mohammad and Gorshkov, Alexey V. and Gong, Zhexuan and Cetina, Marko and Monroe, Christopher},
  journal={Nature},
  volume={623},
  pages={713--717},
  year={2023},
  doi={10.1038/s41586-023-06656-7}
}

@article{katz2025floquet,
  title={Floquet Control of Interactions and Edge States in a Programmable Quantum Simulator},
  author={Katz, Or and Feng, Lei and Porras, Diego and Monroe, Christopher},
  journal={Nature Communications},
  volume={16},
  pages={8815},
  year={2025},
  doi={10.1038/s41467-025-62897-2}
}

@article{karpa2016phasestable,
  title={Phase-Stable Free-Space Optical Lattices for Trapped Ions},
  author={Karpa, Leon and Schmiegelow, Christian Tomas and Kaufmann, Henning and Ruster, Thomas and Schulz, Jonas and Kaushal, Vidyut and Hettrich, Max and Schmidt-Kaler, Ferdinand and Poschinger, Ulrich G.},
  journal={Physical Review Letters},
  volume={116},
  pages={033002},
  year={2016},
  doi={10.1103/PhysRevLett.116.033002}
}

@article{steckmann2506error,
  title={Error mitigation of shot-to-shot fluctuations in analog quantum simulators},
  author={Steckmann, Thomas and Luo, De and Wang, Yu-Xin and Muleady, Sean R and Seif, Alireza and Monroe, Christopher and Gullans, Michael J and Gorshkov, Alexey V and Katz, Or and Schuckert, Alexander},
  year={2025},
  journal={arXiv preprint arXiv:2506.16509},
  doi={10.48550/arXiv.2506.16509}
}

@article{jia2023angle,
  title={Angle-robust two-qubit gates in a linear ion crystal},
  author={Jia, Zhubing and Huang, Shilin and Kang, Mingyu and Sun, Ke and Spivey, Robert F and Kim, Jungsang and Brown, Kenneth R},
  journal={Physical Review A},
  volume={107},
  number={3},
  pages={032617},
  year={2023},
  doi={10.1103/PhysRevA.107.032617}
}

@article{katz2025hybrid,
  title={Hybrid digital-analog protocols for simulating quantum multi-body interactions},
  author={Katz, Or and Schuckert, Alexander and Wang, Tianyi and Crane, Eleanor and Gorshkov, Alexey V and Cetina, Marko},
  journal={arXiv preprint arXiv:2512.21385},
  year={2025},
  doi={10.48550/arXiv.2512.21385}
}

@article{crowley2018quasiperiodic,
  title={Quasiperiodic quantum Ising transitions in 1D},
  author={Crowley, PJD and Chandran, A and Laumann, CR},
  journal={Physical Review Letters},
  volume={120},
  number={17},
  pages={175702},
  year={2018},
  doi={10.1103/PhysRevLett.120.175702}
}

@article{chandran2017localization,
  title={Localization and symmetry breaking in the quantum quasiperiodic ising glass},
  author={Chandran, Anushya and Laumann, CR},
  journal={Physical Review X},
  volume={7},
  number={3},
  pages={031061},
  year={2017},
  doi={10.1103/PhysRevX.7.031061}
}

@article{leibfried2003quantum,
  title={Quantum dynamics of single trapped ions},
  author={Leibfried, Dietrich and Blatt, Rainer and Monroe, Christopher and Wineland, David},
  journal={Reviews of Modern Physics},
  volume={75},
  number={1},
  pages={281},
  year={2003},
  doi={10.1103/RevModPhys.75.281}
}

@article{eckardt2015high,
  title={High-frequency approximation for periodically driven quantum systems from a Floquet-space perspective},
  author={Eckardt, Andr{\'e} and Anisimovas, Egidijus},
  journal={New Journal of Physics},
  volume={17},
  number={9},
  pages={093039},
  year={2015},
  doi={10.1088/1367-2630/17/9/093039}
}

@article{rahav2003effective,
  title={Effective Hamiltonians for periodically driven systems},
  author={Rahav, Saar and Gilary, Ido and Fishman, Shmuel},
  journal={Physical Review A},
  volume={68},
  number={1},
  pages={013820},
  year={2003},
  doi={10.1103/PhysRevA.68.013820}
}

@article{lu2019global,
  title={Global entangling gates on arbitrary ion qubits},
  author={Lu, Yao and Zhang, Shuaining and Zhang, Kuan and Chen, Wentao and Shen, Yangchao and Zhang, Jialiang and Zhang, Jing-Ning and Kim, Kihwan},
  journal={Nature},
  volume={572},
  number={7769},
  pages={363--367},
  year={2019},
  doi={10.1038/s41586-019-1428-4}
}

@article{de2024observation,
  title={Observation of string-breaking dynamics in a quantum simulator},
  author={De, Arinjoy and Lerose, Alessio and Luo, De and Surace, Federica M and Schuckert, Alexander and Bennewitz, Elizabeth R and Ware, Brayden and Morong, William and Collins, Kate S and Davoudi, Zohreh and others},
  journal={arXiv preprint arXiv:2410.13815},
  year={2024},
  doi={10.48550/arXiv.2410.13815}
}

@article{sutherland2021universal,
  title={Universal hybrid quantum computing in trapped ions},
  author={Sutherland, RT and Srinivas, R},
  journal={Physical Review A},
  volume={104},
  number={3},
  pages={032609},
  year={2021},
  doi={10.1103/PhysRevA.104.032609}
}

@misc{SI,
note = "{See Supplemental Material at [URL inserted by publisher] for further details on .... .}"
}
\end{document}